\documentclass[preprint2]{aastex} % demo
\usepackage{graphicx}
\usepackage{amsmath}
\usepackage{natbib}
\usepackage[colorlinks=true,citecolor=blue,breaklinks=true]{hyperref}
\usepackage[switch,pagewise]{lineno}
\usepackage{xspace}
\usepackage{xparse} % used for define optional command 
\usepackage{xcolor} 
\bibpunct{(}{)}{;}{a}{}{,} %% natbib format like A&A and ApJ
\newcommand{\unit}[1]{\ensuremath{\,\mathrm {#1}}}  % text in math mode
\renewcommand{\d}{\ensuremath{\text{d}}}  % differential in formular
\DeclareDocumentCommand\figref{ m g }{{Figure~\ref{#1}\IfNoValueF {#2} {(#2)}}}
\newcommand{\secref}[1]{Section~\ref{#1}}
\renewcommand{\eqref}[1]{Equation~\ref{#1}}

\newcommand{\bvec}[1]{\ensuremath{\boldsymbol{\mathbf{#1}}}}  
\renewcommand{\div}[1]{\nabla\cdot #1} % for divergence
\newcommand{\ars}{AR 11504\xspace}

\shorttitle{stochastic transients}
\shortauthors{Yuan et al.}

\begin{document}

\title{Stochastic transients as a source of quasi-periodic processes in the solar atmosphere}
\author{Ding Yuan\altaffilmark{1,2}}
\email{DYuan2@uclan.ac.uk}
\author{Jiangtao Su\altaffilmark{2}}
\author{Fangran Jiao\altaffilmark{3}}
\author{Robert W. Walsh\altaffilmark{1}}
\altaffiltext{1}{Jeremiah Horrocks Institute,
University of Central Lancashire, Preston, PR1 2HE, United Kingdom}
\altaffiltext{2}{Key Laboratory of Solar Activity, National Astronomical Observatories, Chinese Academy of Sciences, Beijing, 100012, China}
\altaffiltext{3}{Shandong Provincial Key Laboratory of Optical Astronomy and Solar-Terrestrial Environment, Institute of Space Sciences, Shandong University, Weihai, 264209 Shandong, China}
\begin{abstract}
Solar dynamics and turbulence occur at all heights of the solar atmosphere and could be described as stochastic processes. We propose that finite lifetime transients recurring at a certain place could trigger quasi-periodic processes in the associated structures. In this study, we developed a mathematical model for finite lifetime and randomly occurring transients, and found that quasi-periodic processes, with period longer than the time scale of the transients, are detectable intrinsically in form of trains. We simulate their propagation in an empirical solar atmospheric model with chromosphere, transition region and corona. We found that, due to the filtering effect of the chromospheric cavity, only the resonance period of the acoustic resonator is able to propagate to the upper atmosphere, such a scenario is applicable to slow magnetoacoustic waves in sunspots and active regions. If the thermal structure of the atmosphere is less wild and acoustic resonance does not take effect, the long period oscillations could propagate to the upper atmosphere. Such case would be more likely to occur at polar plumes. 
\end{abstract}


\section{Introduction}
\label{sec:intro}
A variety of solar transients and turbulence occurs at the layers close to the visible surface of the Sun: convections \citep{nordlund2009,stein2012}, granulations \citep{rieutord2010}, magnetic reconnections \citep[heatings or flares,][]{hannah2011,cargill2015}, spicular activities \citep{depontieu2007b}, etc. These activities have finite lifetimes and repeat at the same location without nominal periodicities. However, the nature of intermittency in conjunction with finite lifetimes could well lead to a quasi--periodic process in the associated structures.   

In this paper, we aim to propose the idea that stochastic, finite lifetime transients could generate quasi-periodic processes in upper atmosphere of the Sun. This is inspired by the recent discoveries of low--amplitude intermittent transverse oscillations of corona loops \citep{nistico2013,anfinogentov2013,anfinogentov2015}, and the connectivity between spicular activities and quasi-periodic propagating disturbances in coronal holes \citep{jiao2015,samanta2015}. \textbf{The quasi-periodic fast wave trains may also be launched by intermittent impulsive pertubations to reconnection sites \citep{yuan2013,pascoe2013,nistico2014,yang2015}}. We demonstrate the feasibility of this idea by studying how stochastic spicules are connected with propagating disturbances observed in coronal holes and active regions. However, we believe this idea is applicable to other temporal and spatial scales in solar and astrophysics.

Spicules are rapidly evolving elongated transients observed off the solar limb \citep{beckers1968,sterling2000}; while mottles and dynamic fibrils, observed at quiet-Sun and active region plages, respectively, are suggested to be the on-disk counterparts of spicules \citep{hansteen2006,depontieu2007a,rouppevandervoort2009}. Two classes of spicules are identified by \citet{depontieu2007b}, although scepticism remains \citep{sterling2010,zhang2012}. Type I spicules are relatively slowly evolving features, and will eventually fall back due to gravity on a time scale of $3-7\unit{min}$ \citep{depontieu2007b}; Type II spicules will fade off into the background within a lifetime of about $45\unit{s}$ (and a spread about $10-150\unit{s}$) \citep{depontieu2007b}.  

Spicular activity or its on-disk counterpart are believed to be associated with the excessive blue-shifted spectral line emission (upward flows) observed at the footpoints of active regions \citep{delzanna2008,doschek2008,hara2008,tian2011}. Upon observing persistent  upflows and correlated linewidth and intensity variations, \citet{Mcintosh2009},\citet{depontieu2010} and \citet{tian2011} intepreted propagating disturbances observed at footpoints of active region loops as quasi-periodic upflows. However, periodic flows rarely occur in nature, unless they are modulated or waveguided by other thermal or magnetic structures. This idea will be demonstrated in this study. A rival theory is propagating slow mode magnetohydrodynamics (MHD) wave: the propagating disturbances have good periodicy \citep{nakariakov2000,king2003,demoortel2009} and a natural source of sunspot oscillations \citep{yuan2014cf,yuan2014sp,tian2014,su2016a,su2016b}; the intensity (or density) and velocity disturbances oscillate in phase \citep{wang2009a,wang2009b}; the phase speed is temperature-dependent and always smaller than local acoustic speed  \citep{marsh2009,kiddie2012,uritsky2013}; moreover, propagating slow wave could also produce blue-shifted line emission \citep{verwichte2010}, albeit with half wave period. 

Recent simulations suggest upflows will inevitably excite slow waves in active region loops and that a blend of both upflows and slow wave may contribute to the observational features in coronal loops \citep{ofman2012,wang2013}. \citet{fang2015} demonstrates that impulsive heating at the footpoint of a coronal loop will trigger pressure imbalance and excite high-speed upflows. However, as long as the flow become detached with the source, it evolves as a slow wave pulse. \citet{jiao2015} and \citet{samanta2015} found compelling evidence that propagating disturbances at polar plumes have strong correlation with the spicular activities at chromospheric height. Similar dynamics (dynamics fibrils) are observed at active regions \citep{skogsrud2016}. 

In this paper, we propose the idea that quasi-periodicity processes are intrinsic part of stochastic finite lifetime transients, and simulate the propagation of stochastic spicules in the solar atmosphere. \secref{sec:transient} describes the mathematical model of stochastic transients; \secref{sec:numerical} presents the numerical experiment to demonstrate the idea; \secref{sec:observation} compares the synthetic data and Solar Dynamics Observatory (SDO)/Atmospheric Imaging Assembly (AIA) observations at active regions and polar plumes.  

\section{Model of stochastic transients}
\label{sec:transient}
We model a set of sequential transients $g_i(t_i)$ with a peak strength at $t_i$. For simplicity, we assume each transient evolves in an Gaussian profile with an amplitude of $v_i$ and a width of $\sigma_i$: 
\begin{align}
v(t)&=\sum\limits_{i=1}^{N}g_i(t_i)=\sum\limits_{i=1}^{N}v_i \exp\left[\frac{-(t-t_i)^2}{2\sigma_i^2}\right], \label{eq:transients} \\
t_i&=iP_0+\delta t_i, \\
\sigma_i&=\sigma_0+\delta\sigma_i.
\end{align}
The amplitude $v_i$ is an random number that follows a uniform distribution $v_0\mathcal{U}(0,1)$. A transient is assumed to be launched every $P_0$ and shifted by a normally distributed random offset $\delta t_i$, where $\delta t_i/P_1$ has a normal distribution $\mathcal{N}(0,1)$. So the occurrence interval $t_{i+1}-{t_i}$ has an average value of $P_0$ and a standard deviation of $\sqrt{2}P_1$. The width $\sigma_i$ of a transient, relating to its lifetime, has a nominal value of $\sigma_0$ and a random variation $\delta \sigma_i$; and $\delta \sigma_i/\sigma_1$ follows a normal distribution $\mathcal{N}(0,1)$. To reduce the number of free parameters, we fix $P_1=P_0/2$, which allows a small probability that two consecutive transients overlap to some extent. We also set $\sigma_1=\sigma_0/4$, so that $\sigma_i$ is unlikely to be negative. Tests show that even $\sigma_1$ is set to zero or other reasonable values, the spectrum of the time series remains slightly altered. Similar tests were performed on $P_1$, we did not find significant changes to the spectrum. To see the feasibility of this idea, we introduce typical time scales of the dynamics commonly observed in the lower atmosphere of the Sun.

\begin{figure*}[ht]
\centering
\includegraphics[width=0.8\textwidth]{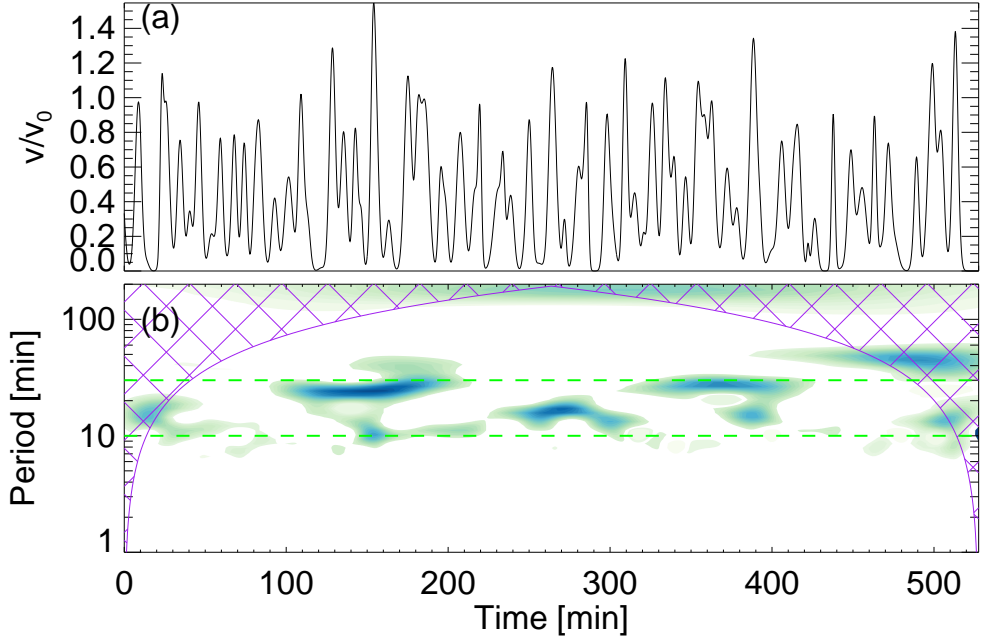}
\caption{(a) Time series of stochastic finite lifetime transient model. (b) Wavelet spectrum illustrates quasi-periodic oscillations at 10-30\unit{min} range, enclosed within the green dashed lines. The Cone-of-Influence is cross-hatched, within which spectrum should be considered as unreliable. We used $P_0=300\unit{s}$ and $\sigma_0=100\unit{s}$. \label{fig:spicule}}
\end{figure*}

Type-I spicules have dynamic time scales at $3-7\unit{min}$ and lifetimes at $2-3\unit{min}$ \citep{depontieu2007b}; Mottles and fibrils are found to have similar time scales \citep{hansteen2006,depontieu2007a,rouppevandervoort2009}. So we used $P_0=300\unit{s}$, and $\sigma_0=100\unit{s}$ to simulate the time series associated by spicular dynamics, see \figref{fig:spicule}{a}. \textbf{The time series is analysed by wavelet transform to show the dynamic spectrum. We applied the Morlet mother function, which is optimised for revealing the oscillatory signals \citep{torrence1998}. In the wavelet spectrum, we only plot the spectral component above the 90\% confidence level. This standard is followed in other wavelet spectra throughout this paper.} Spectral peaks aggregate at about $0.5-1.5\unit{mHz}$ (10--30\unit{min}, \figref{fig:spicule}{b}). Power spectrum of quasi-periodic propagating disturbances observed at coronal holes exhibit periodicities at exactly the same frequency range \citep{mcintosh2010,krishna2011,krishna2012,banerjee2011}. 

\begin{figure*}[ht]
\centering
\includegraphics[width=0.8\textwidth]{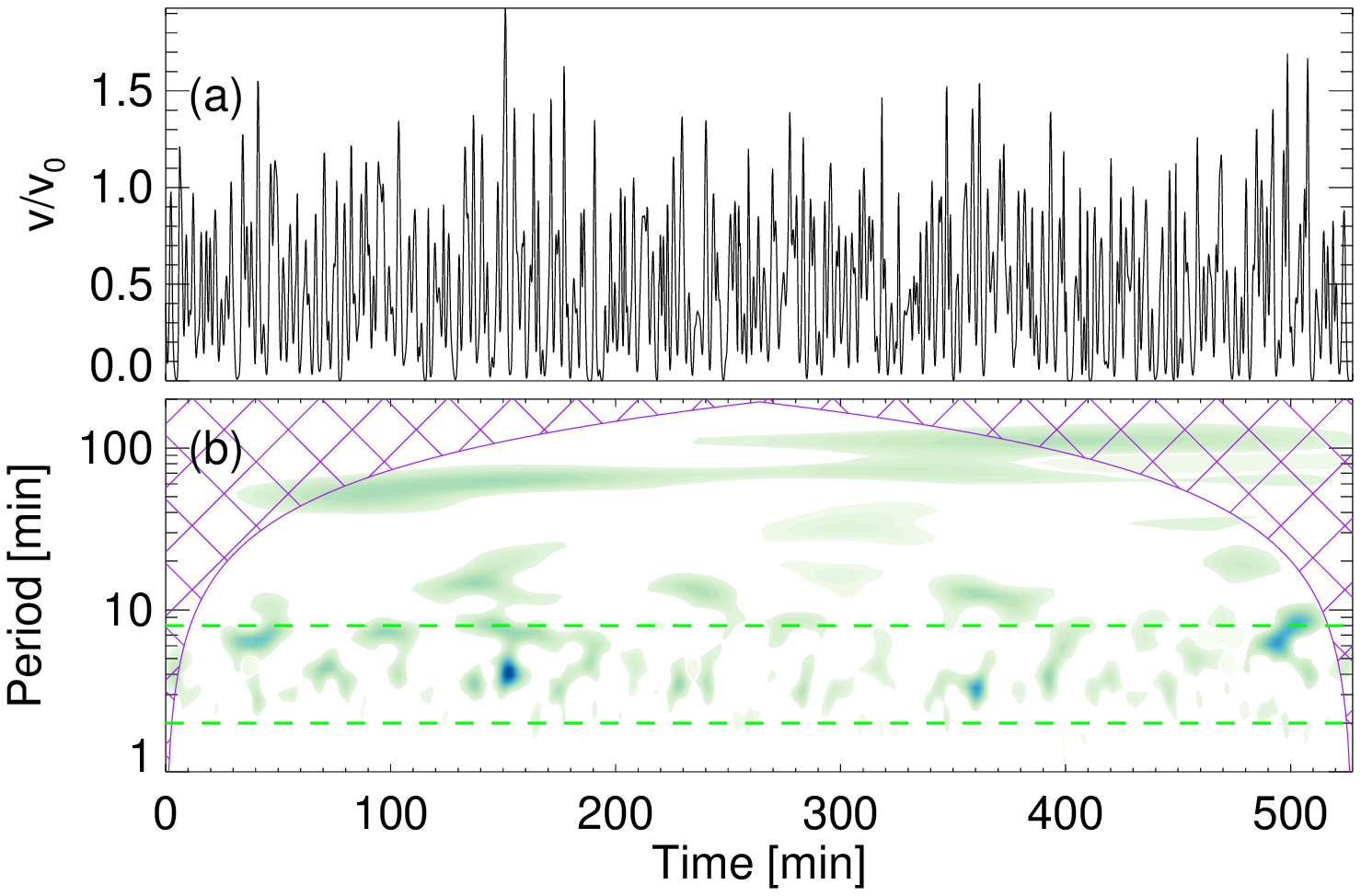}
\caption{Same as \figref{fig:spicule}, but $P_0=80\unit{s}$ and $\sigma_0=30\unit{s}$. Green dashed lines label the prominent quasi-periodic oscillations at 2-8\unit{min} range. The Cone-of-Influence is cross-hatched, within which spectrum should be considered as unreliable. \label{fig:spiculeb}}
\end{figure*}

Type-II spicules recur more frequently (about every $60\unit{s}$) and shorter-lived (about $40\unit{s}$). Shorter time scales are found in nanoflares and high-frequency heatings \citep{porter1984,parker1988,klimchuk2015}. We use $P_0=80\unit{s}$, and $\sigma_0=30\unit{s}$ to model the dynamics of these time scales; prominent periodicities are found at \textbf{2-8 min and 50-70 min bands} (\figref{fig:spiculeb}). Short period propagating disturbances in coronal holes and active regions \citep[e.g.,][]{demoortel2009,banerjee2011,yuan2012sm} and sunspot oscillations \citep[e.g.,][]{yuan2014sp,yuan2014cf,tian2014,su2016} situate in \textbf{2-8 min} period range. 

\section{Numerical experiments}
\label{sec:numerical}
In order to explore how stochastic transients affect the upper atmosphere, we set up a set of numerical experiments, and investigate how the power spectrum is altered observationally throughout its propagation in a stratified atmosphere. We do not attempt to model realistic transients at photospheric temperature and density, but only demonstrate if this sort of perturbations was triggered at the lower atmosphere, what would be the response that could be observed. 

\subsection{Empirical atmosphere model}
\label{sec:empirical}
\begin{figure*}[ht]
\centering
\includegraphics[width=0.8\textwidth]{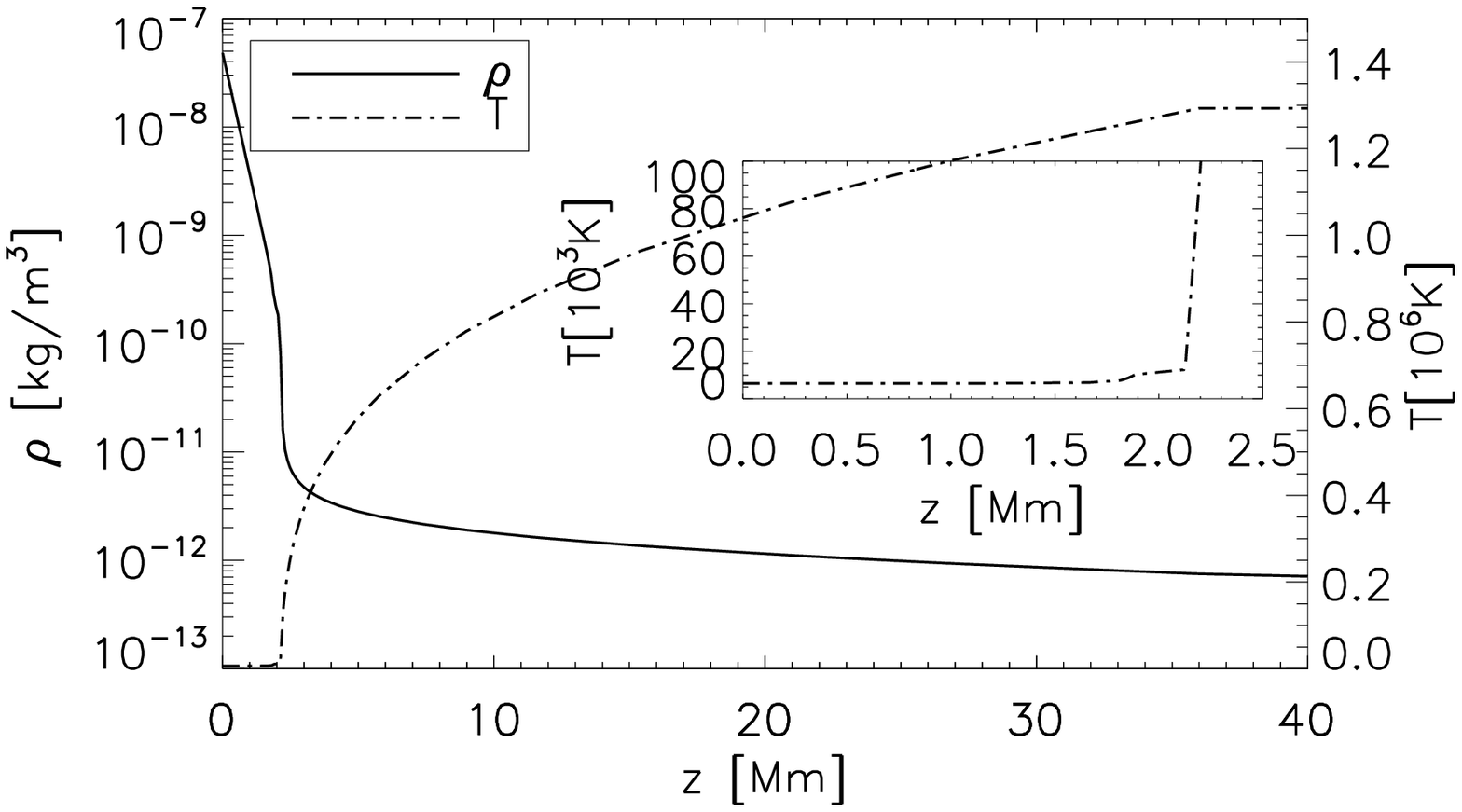}
\caption{Empirical atmosphere model used in this study, the zoomed-in panel plots temperature profile at $z<2.5\unit{Mm}$.\label{fig:spmodel}}
 \end{figure*}

\figref{fig:spmodel} plots the one-dimensional atmosphere model used in this study. Below $z<2.22\unit{Mm}$, we implemented the \citet{maltby1986} mid-age sunspot umbra model; close to the bottom boundary ($z<1.2\unit{Mm}$), we set a constant temperature ($6471\unit{K}$). Above $z=2.22\unit{Mm}$, we adapted the \citet{avrett2008} C7 model; and near the upper boundary ($z\in[36\unit{Mm},40\unit{Mm}]$), we put a constant temperature at 1.3\unit{MK}. This kind of temperature model has been used in a number of previous studies, e.g., \citet{heggland2007,botha2011,snow2015}. 

In hydrostatic equilibrium, the pressure is obtained by integrating the gravity force $\rho g$,
\begin{align}
p(z)&=p(0)\exp\left[\int_0^z-\rho g\d z\right]\\
 p(z)&=\frac{\rho(z)k_\mathrm{B}T(z)}{0.5m_\mathrm{p}},
\end{align}
where $p$, $\rho$ and $T$ are the plasma pressure, density and temperature, respectively, $g=275.43\unit{m\cdot s^{-2}}$ is the gravitational constant, $k_\mathrm{B}$ is the Boltzmann constant, $m_\mathrm{p}$ is the mass of proton. We choose a base density of $\rho(0)=6.6\cdot10^{-8} \unit{kg\cdot m^{-3}}$ at $z=0\unit{Mm}$, so at coronal height the density is of the order of $10^{-13} \unit{kg\cdot m^{-3}}$, see \figref{fig:spmodel}. In one-dimensional MHD simulation, the magnetic field decouples from hydrodynamic equilibrium; we used $B=10\unit{G}$, in line with the $z$-axis, in our simulations.  

The plasma parameters were normalized by choosing a set of three constant: the length $L_0=1.0\unit{Mm}$, the density $\rho_0=8.0\cdot10^{-10}\unit{kg\cdot m^{-3}}$, and the magnetic field strength $B_0=10\unit{G}$. Then the normalization factor for the velocity is $V_0=31.6\unit{km\cdot s}$, the time $t_0=31.7\unit{s}$, the gravitational constant  $g_0=997\unit{m\cdot s^{-2}}$, the temperature $T_0=0.06\unit{MK}$. So the variables become dimensionless by using $z=\hat{z}L_0$, $\rho=\hat{\rho}\rho_0$, $B=\hat{B}B_0$, $t=\hat{t}t_0$, $v=\hat{v}V_0$, $g=\hat{g}g_0$, $p=\hat{p}p_0$, $T=\hat{T}T_0$ etc, where symbols with a hat are dimensionless variables.

\subsection{Magnetohydrodynamic simulation}

We used the AMRVAC package \citep{keppens2012,porth2014} to solve the ideal MHD equations with finite-volume method:

\begin{align}
\frac{\partial\rho}{\partial t}+\div{(\rho\bvec{v})}=&0, \label{eq:mass}\\
\frac{\partial\rho \bvec{v}}{\partial t}+\div\left[\rho\bvec{v}\bvec{v}+\bvec{I}p_\mathrm{tot}-\frac{\bvec{B}\bvec{B}}{\mu_0} \right]=&\rho\bvec{g}, \\
\frac{\partial \epsilon}{\partial t}+\div\left[\bvec{v}(\epsilon+p_\mathrm{tot})-\frac{(\bvec{v}\cdot\bvec{B})\bvec{B}}{\mu_0} \right]=&\rho\bvec{g}\cdot\bvec{v}, \\
\frac{\partial \bvec{B}}{\partial t}+\div(\bvec{v}\bvec{B}-\bvec{B}\bvec{v})=&0,\\
\epsilon=\frac{1}{2}\rho v^2+\frac{p}{\gamma-1}+\frac{B^2}{2\mu_0},\label{eq:state}
\end{align}
where $\bvec{v}$ is the fluid velocity, $\bvec{B}$ the magnetic field vector, $\bvec{g}=-g\bvec{z}$ the solar gravity vector, \textbf{$\bvec{I}$ unit tensor, $\epsilon$ the total energy density, $p_\mathrm{tot}=p+B^2/2\mu_0$ total pressure,} $\gamma$ the adiabatic index, and $\mu_0$ the magnetic permeability of free space.

The AMRVAC code was configured to solve Equations \ref{eq:mass}-\ref{eq:state} in Cartesian coordinates $[x,y,z]$. The physical quantities were assumed to be invariant along the $x$- and $y$-axes (i.e., $\partial/\partial x=\partial/\partial y=0$). We exploited the HLL approximate Riemann solver \citep{harten1983} and implemented the KOREN flux limiter \citep{koren1993}. A three-step Runge-Kutta method was used in time discretization. This configuration has been used in \citet{yuan2015rs} to study propagating fast waves in randomly structured plasmas. 

A solution domain was set at $z\in[0\unit{Mm},40\unit{Mm}]$. At the top boundary, we extrapolated values for the density and pressure; mirroring boundary condition was used for the velocity to minimize reflections. At the bottom boundary we fixed the values of plasma density and pressure, extrapolated from the empirical atmosphere model; the velocity was set up according to the driver model, see \secref{sec:transient}. We used 4000 fixed grid cells without adaptive mesh refinement, so cell size was $\d z=10\unit{km}$. We relaxed the system till $t/t_0=50$, the maximum velocity caused by numerical noise is 0.003 (i.e., $0.1\unit{km\cdot s^{-1}}$). So we could consider the system already in hydrostatic equilibrium, and drivers could be applied at $t/t_0=0$.

\subsection{Single pulse test}

Our model contains an acoustic resonator for slow magnetoacoustic waves: the slow waves get reflected at the sharp density gradient when propagating downwards, and encounter cut-off effect at the upper atmosphere \citep{roberts2006,afanasyev2015}. Theory of the acoustic resonator model could be found in \citet{zhugzhda2008,zhugzhda1983}, and simulations in \citet{botha2011} and \citet{snow2015}.

\begin{figure*}[ht]
\centering
\includegraphics[width=\textwidth]{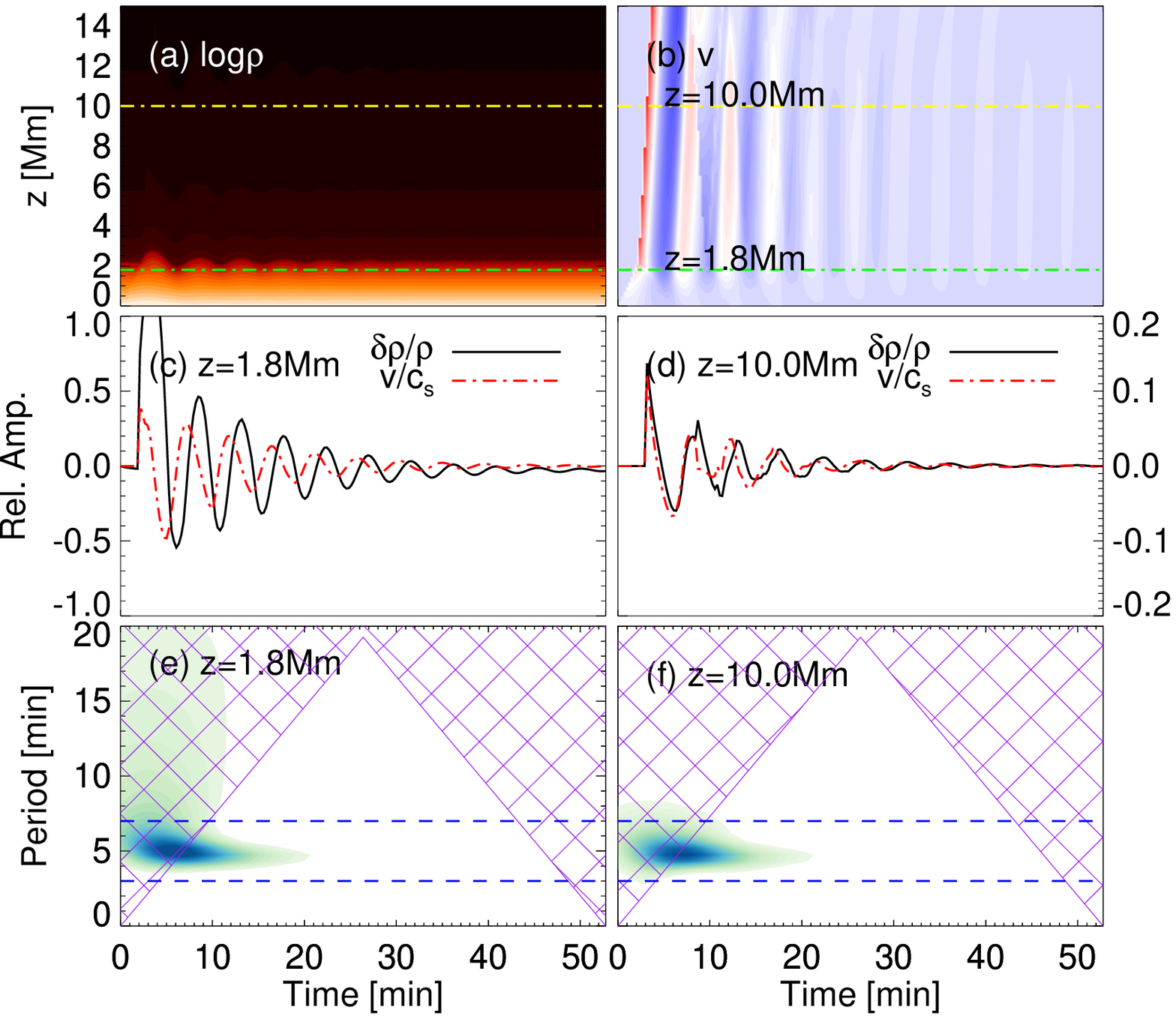}
\caption{(a) and (b) Evolutions of $\log \rho$ and $v$ in the single pulse experiment. Dot-dashed lines mark the positions at $1.8\unit{Mm}$ and $10\unit{Mm}$ where time series are extracted. (c) and (d) Time series of the relative density perturbation $\delta \rho/\rho$ and velocity $v/c_\mathrm{s}$ extracted at $z=1.8\unit{Mm}$ and $z=10.0\unit{Mm}$, respectively, where $c_\mathrm{s} $ is the local sound speed. (e) and (f) Wavelet spectra of the relative density perturbations plotted in (c) and (d), respectively. Blue dashed lines enclose the resonance period range between 3-7\unit{min}. The Cone-of-Influence is cross-hatched, within which spectrum should be considered as unreliable. \label{fig:singlepulse}}
\end{figure*}

A single Gaussian pulse with $\sigma_i=100\unit{s}$ was launched at the bottom boundary. The pulse reached the height at about $2\unit{Mm}$ after about 2 minutes, the average phase speed was about $15\unit{km\cdot s^{-1}}$ (\figref{fig:singlepulse}{a}). Above $2\unit{Mm}$, the density drops dramatically, and shock wave started to form, which is the suggested formation mechanism for chromospheric jets and spicules \citep{heggland2007,depontieu2007b}. The pulse energy did not completely penetrate through the transition region and went off; a significant portion of energy was trapped within the acoustic resonator, bouncing back and forth, see \figref{fig:singlepulse}{c}. The wavelet spectrum of the density perturbation within the resonator $z=1.8\unit{Mm}$ ($T\sim7000\unit{K}$, \figref{fig:singlepulse}{e}) reveals that the pulse initially carried a broadband of signals, and quick involved into a monochromatic oscillation with a period of about $5\unit{min}$. The phase difference between the density and velocity perturbation was initially zero and rapidly split into $\pi/4$ (\figref{fig:singlepulse}{c}), which means a standing slow wave was formed within the resonator \citep{wang2011,yuan2015fm}. At the coronal height, we extracted time series at $z=10\unit{Mm}$ ($T\sim0.8\unit{MK}$). The signal was quasi-periodic (\figref{fig:singlepulse}{f}); and density and velocity oscillated almost in phase (\figref{fig:singlepulse}{d}), meaning propagating slow wave was observed at corona heights \citep[also see][]{wang2009a,wang2009b,yuan2012sm,fang2015}. The average propagating speed was about $100\unit{km\cdot s^{-1}}$. Owing to the lack of persistent energy supply, the slow wave decayed off within a few cycles. 

Our result is consistent with the theory of dispersive evolution of impulsive disturbance in stratified atmosphere \citep{chae2015}. However, we get longer period (or lower resonance frequency) than \citet{botha2011}, this is because we used a less wild temperature profile, thus relative smaller density gradient. It also implies that the resonance frequency is a good probe to the resonator's thermal structure, which may evolve with the age of a sunspot or other solar structures \citep{zhugzhda1983,zhugzhda2008}. We shall note that in the simulations of \citet{heggland2007}, periodic drivers are applied; the effect of the resonator is barely seen. This may be owe to the fact that they aim at simulating shock waves with large amplitudes and thus render the trapped energy less prominent. However, irregularities of the shock fronts could be seen in Figure 11 of \citet{heggland2007}, which may be the signature of the trapped energy. 

\subsection{Perturbation by stochastic transients}
\begin{figure*}[ht]
\centering
\includegraphics[width=\textwidth]{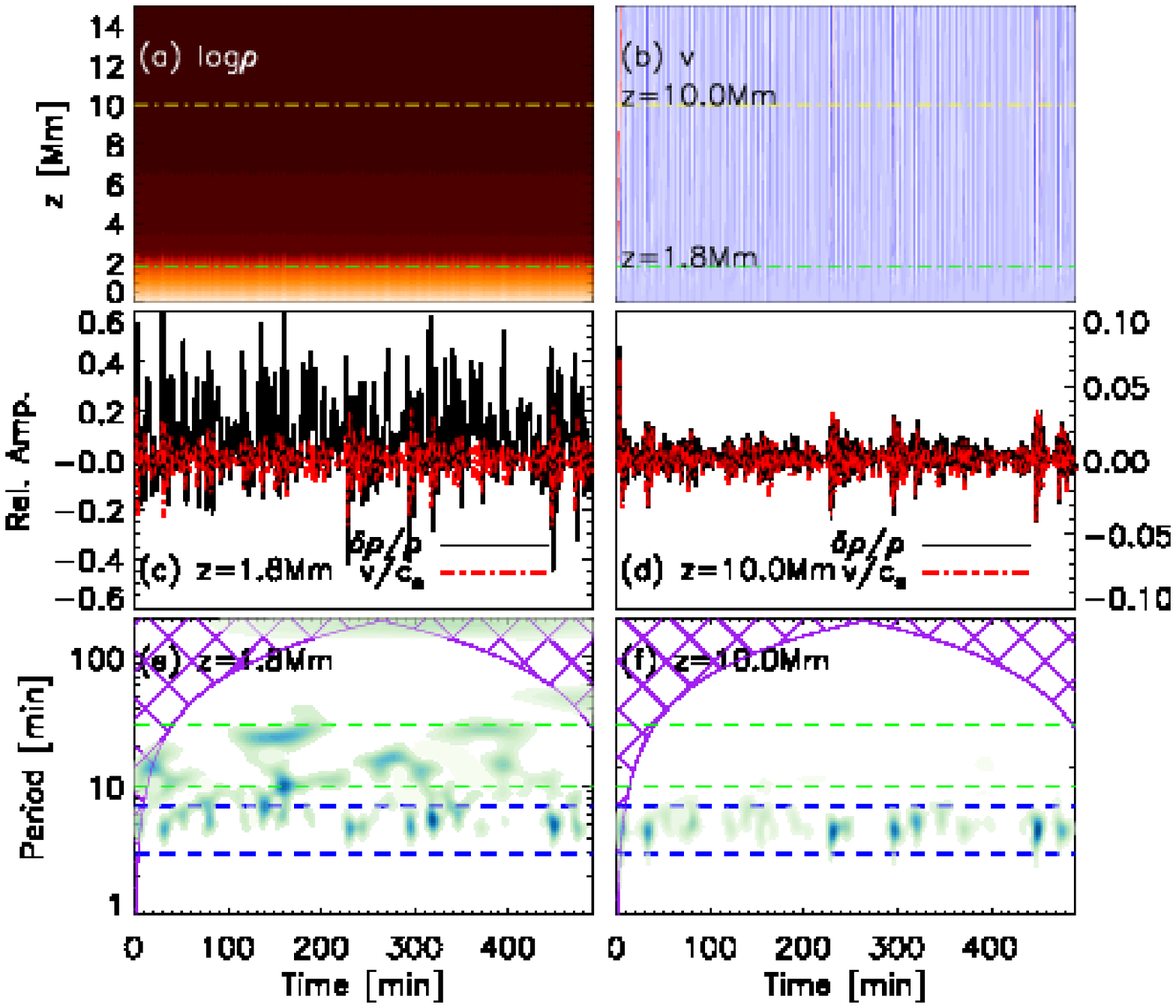}
\caption{Same as \figref{fig:singlepulse}, but with a stochastic driver. Two green dashed lines labels the quasi-periodic long period oscillations brought in by the stochastic driver as illustrated in \figref{fig:spicule}, whereas two blue dashed lines enclose the resonance period range, as shown in \figref{fig:singlepulse}. The Cone-of-Influence is cross-hatched, within which spectrum should be considered as unreliable. \label{fig:random_lo}}
\end{figure*}

After the single pulse experiment, we applied the stochastic transients model (\eqref{eq:transients}), times series and parameters are illustrated in \figref{fig:spicule}. \figref{fig:random_lo} presents time distance plots of density and velocity, and the time series and relevant wavelet analysis. Shock waves are formed at a height where density drops dramatically (\figref{fig:random_lo}{a}), the amplitude could be as large as 60\% of the background density (\figref{fig:random_lo}{c}). At $z=1.8\unit{Mm}$, the phase between density and velocity perturbations have mixed feature, while at $z=10\unit{Mm}$, it is almost zero, indicating that propagating slow waves are found at the coronal height. The wavelet spectrum reveals that at $z=1.8$, long period oscillations ($10-30\unit{min}$) and the resonance period ($\sim5\unit{min}$) co-exist within the acoustic resonator (\figref{fig:random_lo}{e}); while only the resonance period is allowed to leak from chromosphere (\figref{fig:random_lo}{f}). This is consistent with \citet{snow2015}, who demonstrated that even long period oscillations are implemented in the driver, their power would be 2-3 orders of magnitude smaller than the resonance period. Long period oscillations are indeed observed at active regions and other coronal structures \citep{wang2009b,marsh2009,yuan2011lp}; this effect is ascribed to modification to the cut-off period of slow wave by inclined magnetic field, see \citet{bel1977,mcintosh2006,jess2013,yuan2014cf,afanasyev2015}.

We also notice that the resonance within the acoustic cavity and the leakage to the corona are not persistent, but forms a series of wave trains. This is a novel feature that no other study could be able to simulate, and it is consistent with the fact that propagating disturbances at active region loops are detected in forms of wave trains \citep{demoortel2009,demoortel2012}. Moreover, the mixture of both long period and short period oscillations in the acoustic resonator is remarkably similar to the sunspot oscillations observed at chromospheric height using the Nobeyama RadioHeliograph \citep{chorley2010,chorley2011}.

Long period oscillations are also detected at coronal plumes \citep{ofman1999,gupta2010,krishnaprasad2011,jiao2015}, At low corona, the magnetic field pressure dominates over gas pressure, a plume could be simply assumed to be isothermal \citep{delzanna1998}. In this study, we did not use an isothermal model, but simply apply the driver at $z=2.3\unit{Mm}$, and demonstrate that altering the thermal structure of the solar atmosphere could also lead to the leakage of long period oscillations.

\begin{figure*}[ht]
\centering
\includegraphics[width=\textwidth]{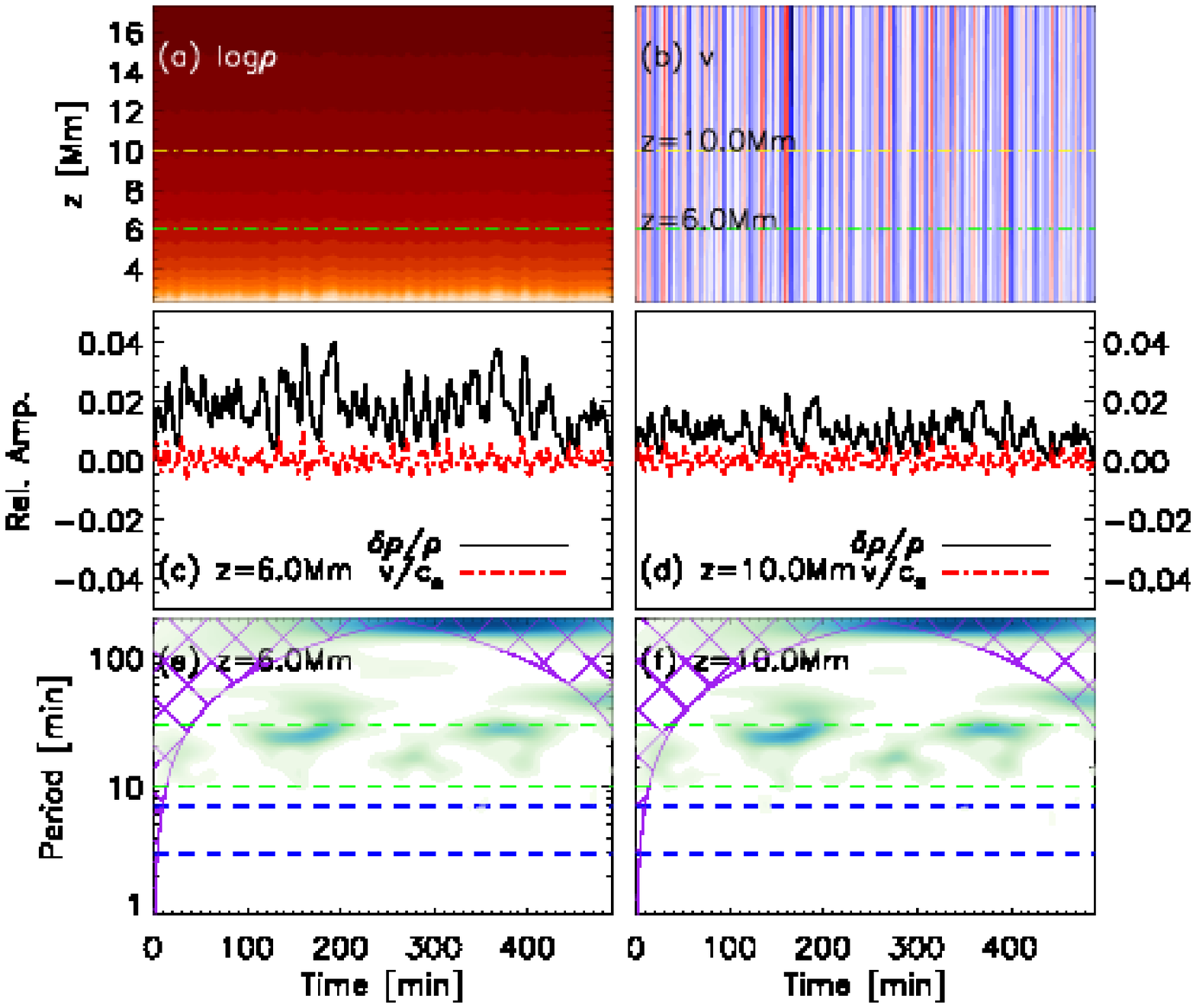}
\caption{Same as \figref{fig:random_lo}, but with the driver implemented at $z=2.3\unit{Mm}$ and time series were extracted at $z=6\unit{Mm}$ and $z=10\unit{Mm}$, respectively. Two green dashed lines labels the quasi-periodic long period oscillations brought in by the stochastic driver as illustrated in \figref{fig:spicule}, whereas two blue dashed lines enclose the resonance period range, as shown in \figref{fig:singlepulse}. The Cone-of-Influence is cross-hatched, within which spectrum should be considered as unreliable.\label{fig:random_hi}}
\end{figure*}

\figref{fig:random_hi} presents the result if we apply the driver at $z=2.3\unit{Mm}$ ($T=0.1\unit{MK}$); this scenario skip the chromospheric cavity. In contrast to being launched deeper, the density and velocity perturbations do not form shock waves, but stay at small amplitude level (\figref{fig:random_hi}(c) and (d)). The long period wave trains manage to propagate into coronal heights.  

\section{Observations in the solar atmosphere}
\label{sec:observation}

\subsection{Oscillations in sunspot and active region loops}

\begin{figure*}[ht]
\centering
\includegraphics[width=\textwidth]{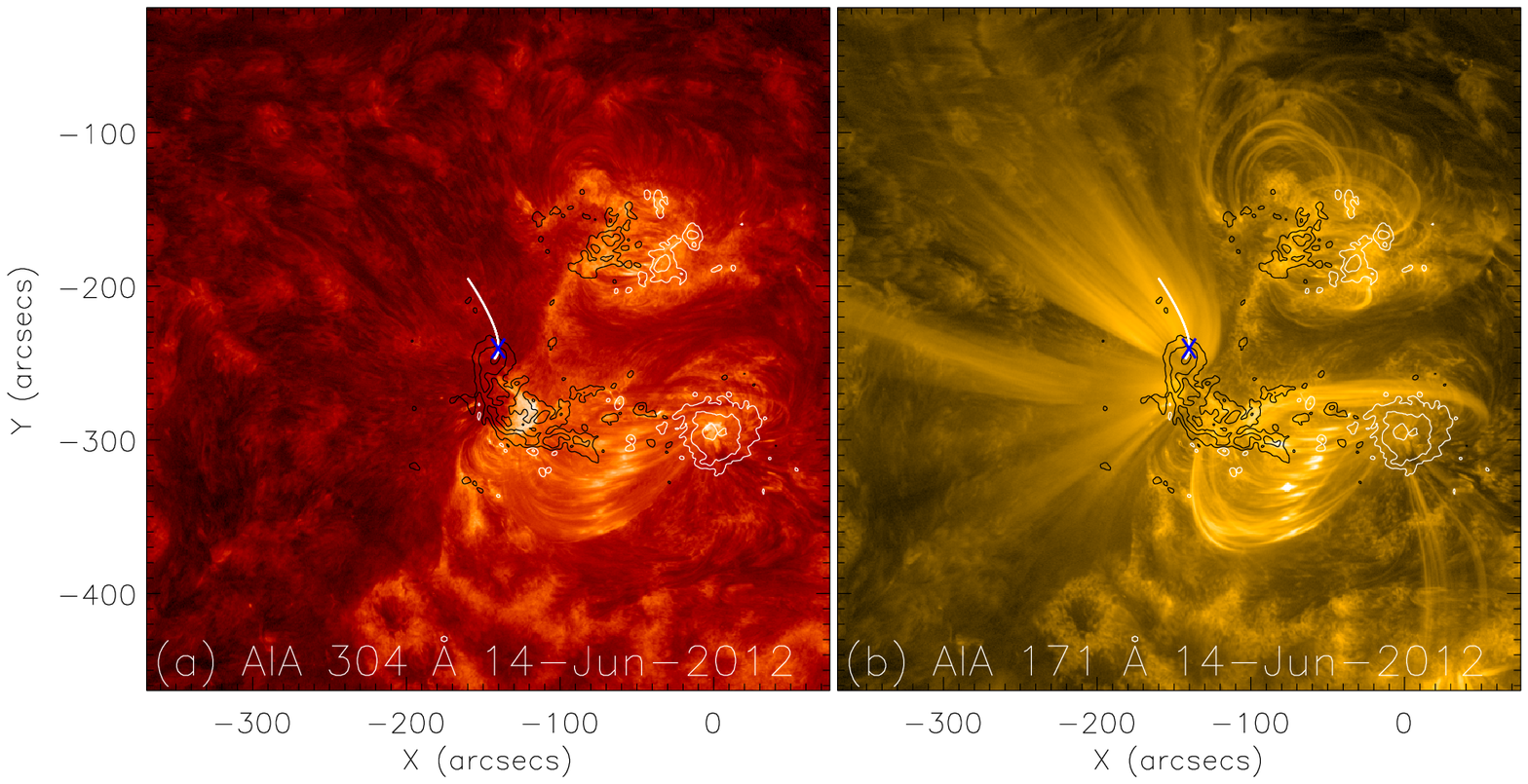}
\caption{Field-of-view (FOV) of \ars observed by the SDO/AIA 304 \AA{} (a) and 171 \AA{} (b) channels on 14 Jun 2012. The Helioseismic and Magnetic Imager (HMI) Line-of-sight magnetogram is contoured at $\pm1800\unit{G},\pm900\unit{G},\pm300\unit{G}$. White and black contours are north and south polarities, respectively. The white slice follows a coronal loop, and the blue cross labels the positions where time series are extracted for wavelet analysis in \figref{fig:sunspot}.
\label{fig:fov_sp}}
\end{figure*}

\begin{figure*}[ht]
\centering
\includegraphics[width=0.8\textwidth]{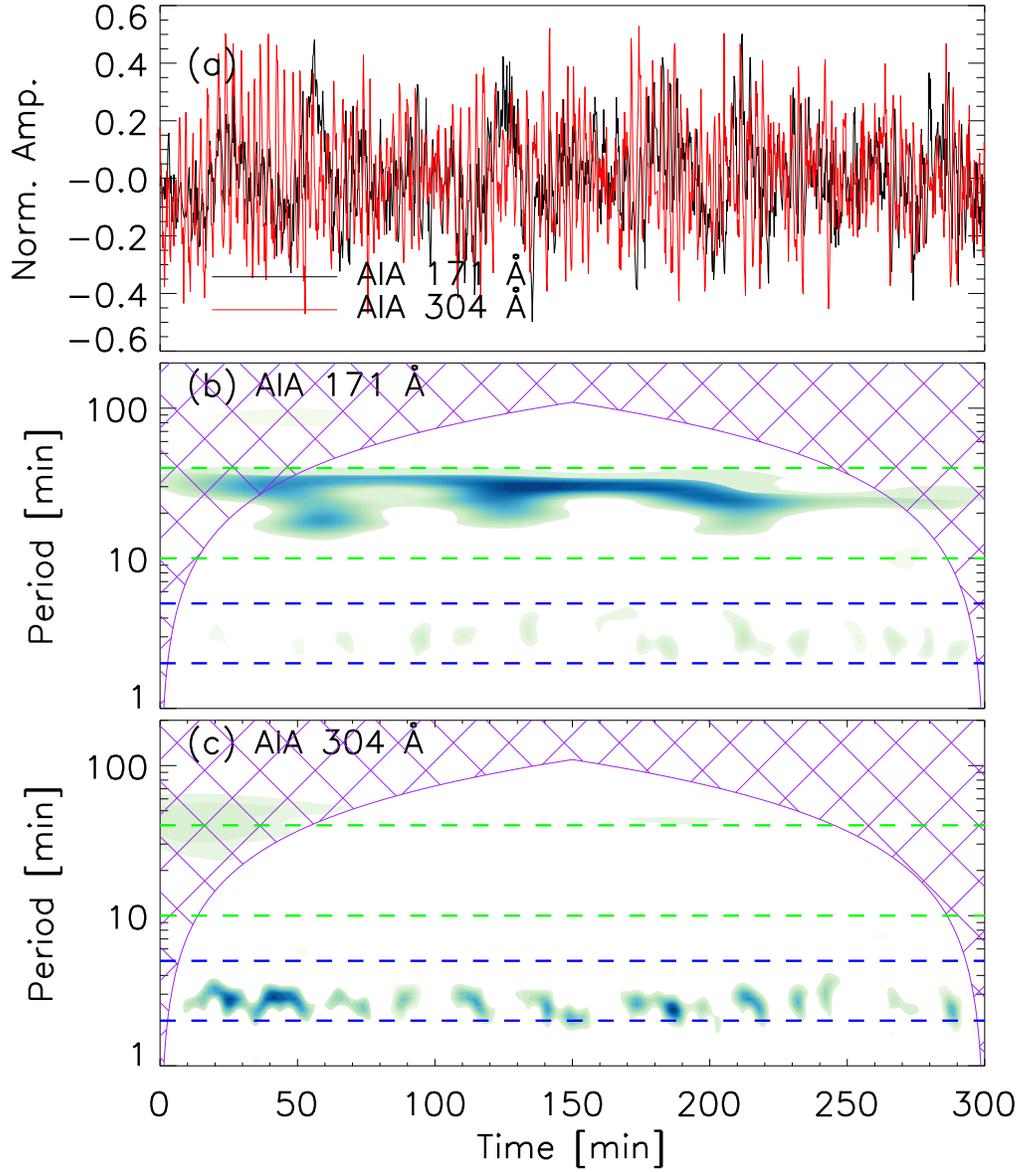}
\caption{(a) Time series of relative emission intensity extracted at a coronal loop marked in \figref{fig:fov_sp}. The peak-to-peak amplitude is normalized to $\pm0.5$. (b) and (c) are the wavelet spectra for the AIA 171 \AA{} and 304 \AA{}, respectively. The start time is 16:00 UT 14 Jun 2012. Two green dashed lines labels the quasi-periodic long period oscillations at 10-30\unit{min}, whereas two blue dashed lines enclose oscillations at 3\unit{min} band. The Cone-of-Influence is cross-hatched, within which spectrum should be considered as unreliable. \label{fig:sunspot}}
\end{figure*}

In the solar atmosphere, thermal structure of a sunspot and the associated active region (AR) resembles the used empirical atmospheric model (see \secref{sec:empirical}). Therefore, we selected a 5-hours AIA observation on \ars (\figref{fig:fov_sp}). The start time is 16:00:00 UT 14 June 2012. We used the AIA 304 \AA{} and 171 \AA{} channels, which have nominal response temperatures at $85,000\unit{K}$ (\ion{He}{2}) and $850,000\unit{K}$ (\ion{Fe}{9}),  respectively \citep{boerner2012}. This AR has already been studied intensively in \citet{su2013}. A coronal loop anchoring at the trailing spot of \ars was tracked (\figref{fig:fov_sp}). We extracted time series of the emission intensity in AIA 304 \AA{} and 171 \AA{}, removed the trend of a $30\unit{min}$ running average, and normalized the amplitude to $\pm0.5$. \figref{fig:sunspot} plots the time series and the relevant wavelet analysis. \textbf{Although we cannot guarantee that two time series were taken at two different heights of the same loop, but the oscillations at diffuse coronal structures have a correlation area of about 10 Mm long and a few Mm wide \citep{demoortel2002a}, within which the wavelet spectrum is barely altered. So the ambiguity owe to the projection effect will not affect our conclusions.}

We could see that in both AIA 304 \AA{} and 171 \AA{} channels, one could detect prominent oscillations at three minute in form of wave trains (\figref{fig:sunspot}), which has been reported in a number of studies \citep[e.g.,][]{demoortel2002,king2003,yuan2014cf}. In the meanwhile, we also detect significant long period oscillations at $10-30\unit{min}$ at both heights. The leakage of long period oscillations was studied previous, e.g.,\citet{yuan2011lp}, \citet{marsh2009}, \citet{wang2009b}. The mechanism for the leakage is not fully understood: the inclined magnetic field could lower cut-off frequency \citep{depontieu2005,mcintosh2006,yuan2014cf}, evanescent slow waves with long penetration depths re-gain propagating wave feature at coronal heights \citep{fleck1991,yuan2011lp}, or long period oscillations are somehow generated at coronal height.   

\begin{figure*}[ht]
\centering
\includegraphics[width=0.8\textwidth]{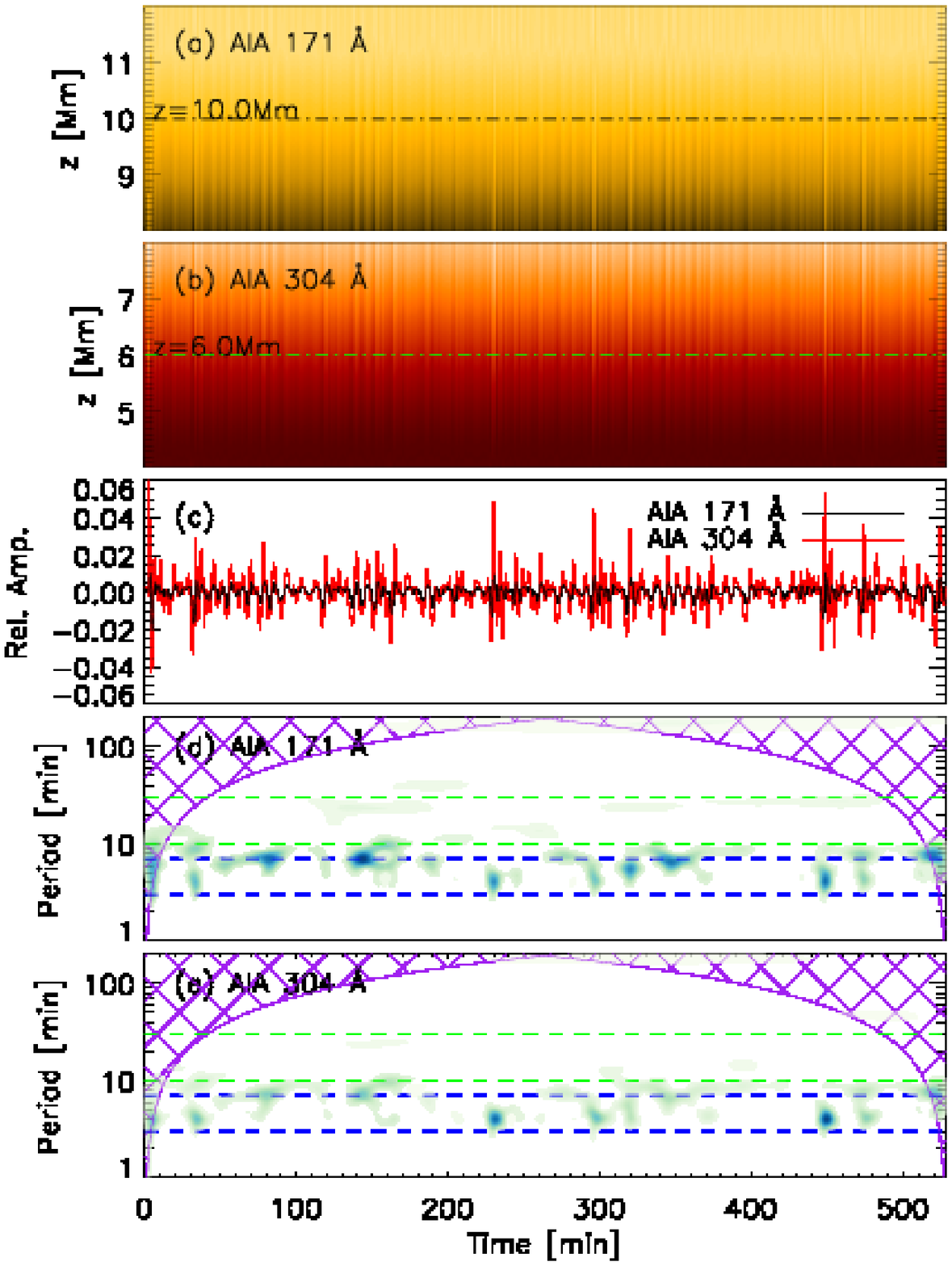}
\caption{(a) and (b) Evolution of the synthetic AIA 171 \AA{} and 304 \AA{} emission intensities for the simulation with driver applied at $z=0\unit{Mm}$. The black dot-dashed line labels the height at $z=10\unit{Mm}$, where a time series is extracted and analysed by wavelet transform; The green dot-dashed line denotes the height at $z=6\unit{Mm}$ where the 304 \AA{} counterpart is obtained: time series and wavelet power spectrum. (c) Relative intensity variations of 304 \AA{} extracted at $z=6\unit{Mm}$ and 171 \AA{} at $z=10\unit{Mm}$, respectively. (d) and (f) The wavelet power spectrum for the relative intensity variation of 304 \AA{} and 171 \AA{}, respectively. The Cone-of-Influence is cross-hatched, within which spectrum should be considered as unreliable. Two green dashed lines labels the quasi-periodic long period oscillations brought in by the stochastic driver as illustrated in \figref{fig:spicule}, whereas two blue dashed lines enclose the resonance period range, as shown in \figref{fig:singlepulse}.  \label{fig:random_lo_aia}}
\end{figure*}

To compare with the observational data, we synthesized the AIA 304 \AA{} and 171 \AA{} emission intensities \citep[details of forward modelling is available at][]{yuan2015fm,fang2015}, by using the CHIANTI atomic emission database \citep{dere1997,landi2013}. \figref{fig:random_lo_aia} presents the time-distance plots, the time series extracted at $z=6\unit{Mm}$ (304 \AA{}) and $10\unit{Mm}$ (171 \AA{}) and the associated wavelet analysis. The time series for the synthetic AIA 304 \AA{} and 171 \AA{} emission intensities exhibit only oscillations at the resonance period ($\sim5\unit{min}$); while long period oscillations are occasionally detected at low confidence level. In our simulation, the magnetic field line is vertical; the magnetoacoustic cut-off frequency is not modified at all. In an active region, magnetic field are bent to connect another polarity, so significant long period oscillations are more likely to be detected \citep{yuan2011lp,yuan2014cf}.

\subsection{Propagating disturbances in polar plumes}
\begin{figure*}[ht]
\centering
\includegraphics[width=\textwidth]{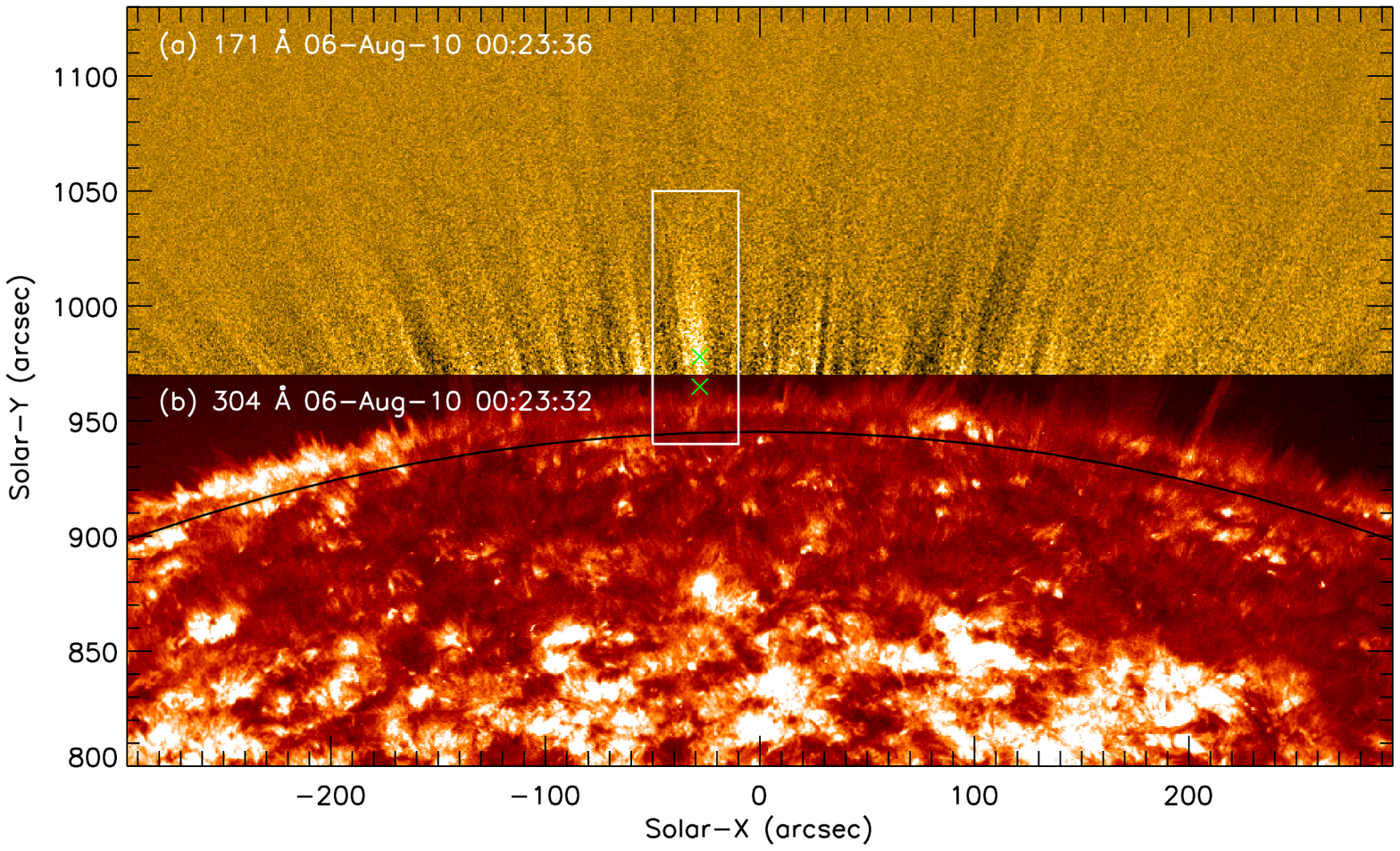}
\caption{A northern coronal hole observed by the AIA 171 \AA{} (a) and 304 \AA{} (b) channels on 06 August 2010. A white rectangle marks the plume of interest: its base is observed by the 304 \AA{} channel, while the upper part is visible in 171 \AA{}. Two time series were extracted at both channels (green crosses) and analysed in \figref{fig:plume}. \label{fig:fov_ch}}
\end{figure*}

\begin{figure*}[ht]
\centering
\includegraphics[width=\textwidth]{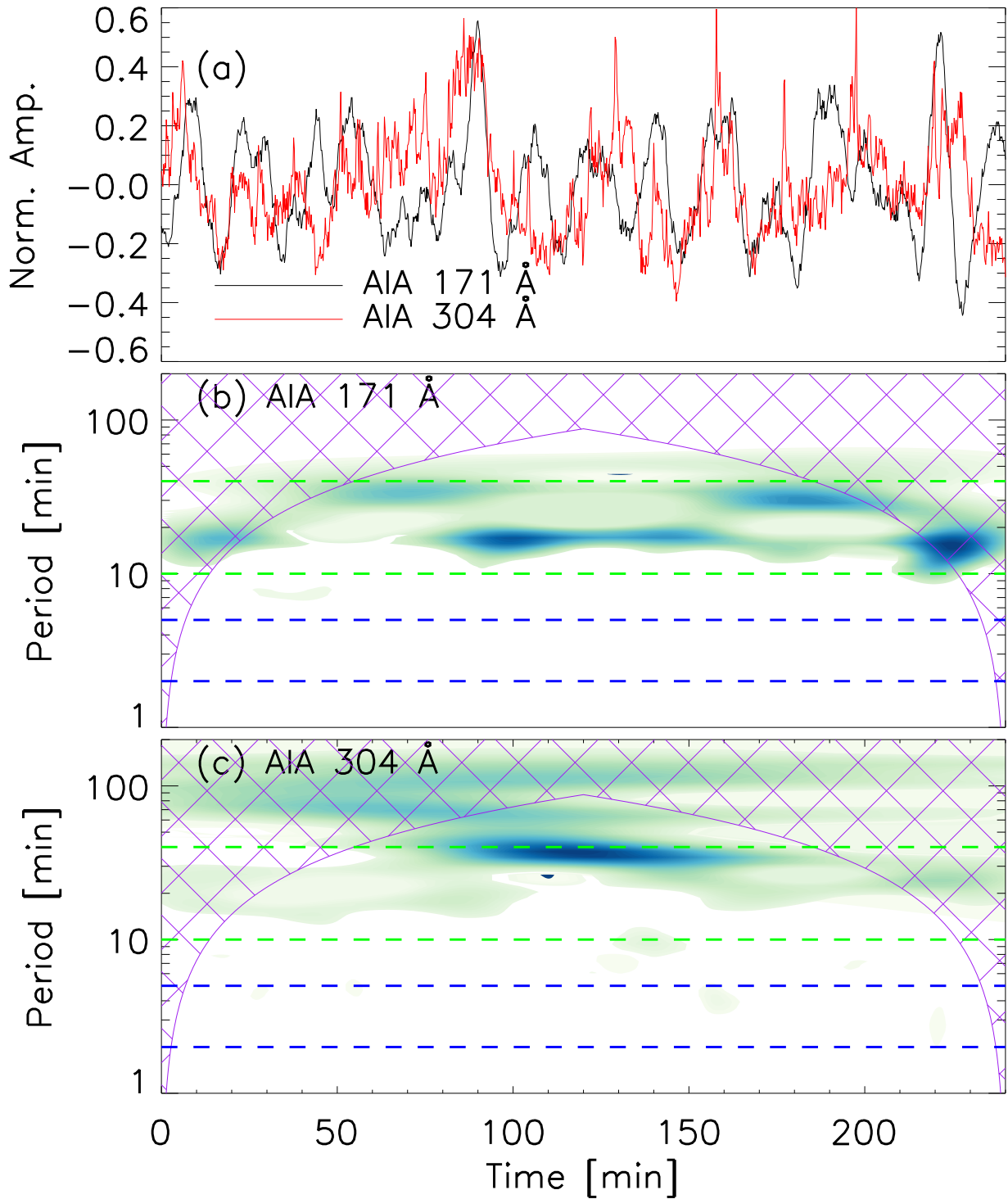}
\caption{Same analysis as in \figref{fig:sunspot}, but for the time series extracted at a plume as labeled in \figref{fig:fov_ch}. Two green dashed lines label the quasi-periodic long period oscillations at 10-30\unit{min}, whereas two blue dashed lines enclose oscillations at 3\unit{min} band. The Cone-of-Influence is cross-hatched, within which spectrum should be considered as unreliable. \label{fig:plume}}
\end{figure*}

To investigate the oscillations in polar plumes, we select a set of 4-hours AIA observations on a coronal hole, starting from 21:50 UT 5 August 2010. A few polar plumes are identifiable at low corona in the AIA 171 \AA{} channel, and their bases could be well traced in the AIA 304 \AA{} channel (\figref{fig:fov_ch}). The plume structure enclosed in rectangle are chosen for further analysis. We extracted the average emission intensity within the small boxes in the 304 \AA{} and 171 \AA{} FOVs, respectively. We used the original light curves of 304 \AA{} bandpass, while for the 171 \AA{} bandpass, we removed a trend of the $30\unit{min}$ moving average and smoothed the time series. The amplitudes are normalized to $[\pm0.5]$. Long period oscillations at $10-30\unit{min}$ are apparent in the light curve and wavelet spectrum of 304 \AA{} and 171 \AA{} (\figref{fig:plume}). 

\begin{figure*}[ht]
\centering
\includegraphics[width=0.8\textwidth]{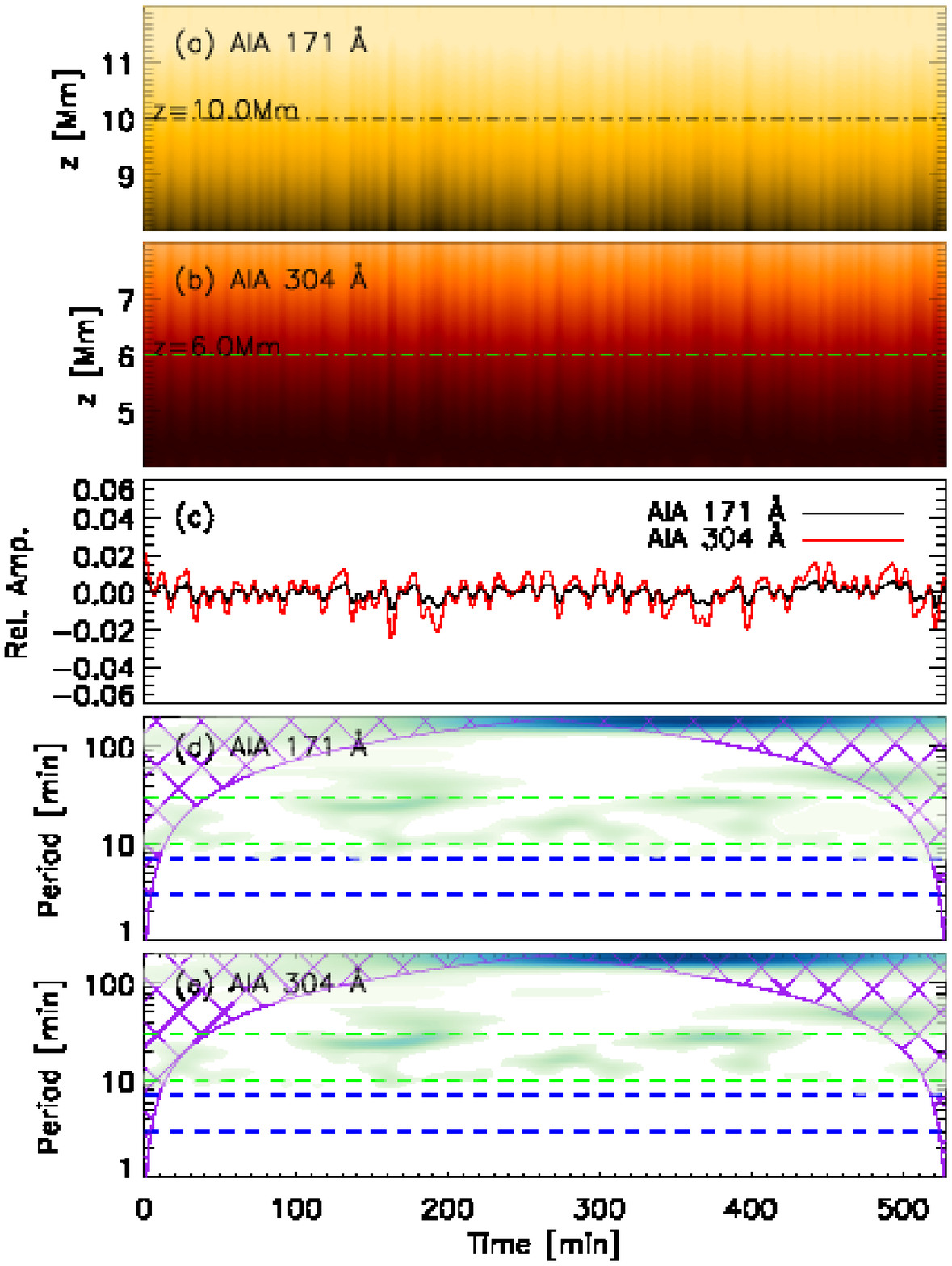}
\caption{Same as \figref{fig:random_lo_aia}, but for the simulation with driver applied at $z=2.3\unit{Mm}$.  The Cone-of-Influence is cross-hatched, within which spectrum should be considered as unreliable. Two green dashed lines labels the quasi-periodic long period oscillations brought in by the stochastic driver as illustrated in \figref{fig:spicule}, whereas two blue dashed lines enclose the resonance period range, as shown in \figref{fig:singlepulse}. \label{fig:random_hi_aia}}
\end{figure*}

Similar procedures are implemented to the simulation with driver launched at $z=2.3$. \figref{fig:random_hi_aia} illustrates the forward modelling result to compare with oscillations in polar plumes. \textbf{Indeed, the long period oscillations propagate to coronal height, and are detectable in the UV/EUV bandpasses.} 

\textbf{In the simulation, we only use the time scale of spicular activities that are found to be highly correlated with quasi-periodic propagating disturbances in polar plumes \citep{jiao2015,samanta2015}. However, as \citet{krishna2012} shows multiple periods co-exist in the propagating disturbances, therefore other time scale are needed to fully address this issue.
}

\section{Conclusions}
In this study, we propose that stochastic, finite-life time transients could generate quasi-periodic processes in the solar atmosphere; and we suggest an mathematical model of stochastic transients in form of Gaussian profiles. Then we use the typical time scales of spicular activities and simulate the propagation of stochastic transients in an empirical atmosphere model. The existence of an chromospheric resonator filters out the long period oscillations, only the resonance period is able to propagate to the upper atmosphere. Observations with SDO/AIA 304 \AA{} and 171 \AA{} bandpasses, which are sensitive to the chromospheric and coronal heights, produce consistent results. We also investigate the case that the thermal structure are changed into a smoother form and the acoustic resonator does not exist. Such a scenario is applicable to polar plumes. Indeed, both simulations and observations clearly measures significant long period oscillations. 

Our model is applicable to many field in geophysics, solar, stellar and astrophysics. The time scale could scaled to any range of interest, we only provide a possible application. 

\acknowledgements
The research was supported by the Open Research Program KLSA201504 of Key Laboratory of Solar Activity of National Astronomical Observatories of China (D.Y.).

\bibliographystyle{apj}
\bibliography{yuan2016st}
\end{document}